\documentclass[pra,twocolumn,showpacs,preprintnumbers,amsmath,amssymb,revtex4-1]{revtex4-1}
\usepackage{amsfonts, amsmath, bm, bbm, graphicx, epsfig, epstopdf}
\usepackage{dcolumn}
\usepackage{textcomp}
 \usepackage[caption=false]{subfig}
 \usepackage[caption=false]{subfig}
\usepackage{soul}
    \usepackage{braket}
\renewcommand\bra[1]{{\langle{#1}|}}
\makeatletter
\renewcommand\ket[1]{%
  \@ifnextchar\bra{\k@t{#1}\!}{\k@t{#1}}%
}
\newcommand\k@t[1]{{|{#1}\rangle}}
\makeatother

\usepackage{subfig}

\begin{document}

\bibliographystyle{apsrev}

\title{Optimal Control for Rydberg multi-qubit operations}
\author{Hossein Abedi$^1$, Mohammadsadegh Khazali$^1$,  Klaus M\o lmer$^2$}
\affiliation{$^1$ Dept of Physics, University of Tehran, Tehran, Iran\\
$^2$ Niels Bohr Institute, Blegdamsvej 17, 2100 Copenhagen, Denmark}

\date{\today}
\begin{abstract}
Quantum computing algorithms can be decomposed into a universal set of elementary one- and two-qubit gates. Different physical implementations of quantum computing, however, employ interactions that permit direct conditional dynamics on multiple qubits in a single step. In this work, we leverage quantum optimal control techniques to design single continuous laser pulses that implement multi-qubit controlled-phase, -NOT and -swap (Fredkin) gates on Rydberg atom quantum processors.  The identification of robust multi-qubit operations leads to reduced operation time and less decoherence, and the control field provides continuous protection of the atoms from environmental noise. Notably, we find that the controlled-swap (Fredkin) gate, implemented using this approach achieves 99.74\% fidelity while accounting for imperfections such as spontaneous emission, laser fluctuations, and Doppler dephasing.
\end{abstract}

\maketitle

\section*{introduction}

 One of the significant challenges in constructing a full-scale quantum computer is the substantial number of basic gates needed, even for small quantum circuits. The complexity and cost of a quantum circuit are typically quantified by the number of controlled-NOT (C-NOT) gates it employs. Various strategies have been employed to reduce the number of C-NOT gates required in a given circuit \cite{Bar95, Pai94, Dal07, Khan01, Bul04}. 
 Implementing the circuit model in atomic quantum processors often necessitates precise single-site addressing of individual atoms with numerous pulsed lasers. These complexities can be mitigated by using quantum optimal control (QOC) techniques to design a single, continuous, and in some cases global pulse that accomplishes the same intricate task \cite{Goerz2019,Koch2022,Pal02,Tes01,Tre06, Sch09, Mul11, Egg14, Sch11, Geo14,Cro15}.

Controlled-phase operations are native to Rydberg platforms, making them a natural choice for quantum gate implementation. Consequently, previous efforts in quantum optimal control for Rydberg systems have primarily focused on optimizing controlled-phase operation types \cite{Eve23,Sve23}. In contrast, controlled-swap (C-SWAP) operations are not inherently native to Rydberg systems. Previous attempts to realize such gates using the anti-blockade mechanism have encountered significant challenges due to their extreme sensitivity to experimental imperfections, including interatomic distance fluctuations and Doppler shifts.  

This article proposes an alternative approach to implementing C-SWAP gates by leveraging the intrinsic Rydberg blockade mechanism. Since population rotations in the collective qubit basis are governed by quantum interference effects and controlled by collective Rydberg excitations, the transition rates depend on the initial qubit configuration. By employing quantum optimal control techniques, we tailor the driving fields to exploit these qubit-dependent transition rates, steering the system toward the desired output. This approach enhances the feasibility of C-SWAP gates in Rydberg platforms while mitigating common experimental limitations.

To generate a general optimal control Rydberg multi-qubit operation, we need two types of laser drivings. 
 To ensure that the operation depends on the states of multiple atoms, all interacting atoms must be laser-excited from a specific qubit state to a strongly interacting Rydberg state. This arrangement would be sufficient for controlled phase operations. To directly implement controlled-rotation or -swap gates, we consider an additional laser transition between qubit basis states for atoms undergoing rotation.
Next, quantum optimal control (QOC) techniques are applied to optimize the intensity, phase, and frequency of the laser beams to achieve the desired operation outcome across all input qubit configurations. To optimize the pulse shapes for our system, we employ the Krotov method, originally developed for the optimal attitude maneuver of space vehicles. 
This approach is demonstrated in the present work by implementing two prominent universal multi-qubit gates used in quantum algorithms.  
In the main text, we study a complex three-qubit controlled-swap operation as a general example, while a less intricate C$_k$-Z operation is discussed in the supplementary material.

The Fredkin gate is a fundamental multi-qubit gate that, when combined with the Hadamard gate, forms a universal set for quantum computation. It plays a crucial role in various quantum algorithms, including quantum fingerprinting \cite{Buh01}, quantum state preparation \cite{Oza14}, quantum state estimation \cite{Eke02}, and optimal cloning \cite{Hof12}.
Implementing a three-qubit Fredkin gate with a circuit model approach requires 8 CNOT and 9 single-qubit gates \cite{Kim18}. This would require 49 laser pulses in a Rydberg quantum processor. Here we implement the same gate in one step using quantum optimal control. Since decoherence and implementation imperfections add up with the number of operations, the proposed scheme significantly improves the fidelity of the gate.

In the Rydberg processors, there are two competing sources of infidelity, spontaneous emission and rotation errors \cite{Kha24,KhaFermi23,Kha23,Kha20,Kha21IJAP,Kha19,Kha15,Mom25,Khaz25}. When tuning the intensity of the laser, a large Rabi frequency reduces the operation time and hence the spontaneous emission while increasing the population rotation errors. Here, we show that the quantum optimal control can steer the controlling parameters in a way to suppress the population of short-lived Rydberg states and simultaneously ensure perfect population rotation leading to high-fidelity operation.

Another advantage of using a single pulse for operations is the continuous laser protection of the Rydberg atoms \cite{The16,Kha23}.
A commonly used method for Rydberg quantum gates is the $\pi$-gap-$\pi$ scheme, where control qubits are left in Rydberg states for a gap period while target qubits undergo laser rotations. 
However, for highly excited Rydberg atoms with a large principal quantum number (e.g.,  $n=100$), the subsequent de-exciting $\pi$ pulse fails to fully return the Rydberg population to the qubit basis \cite{Mal15, Gra19}.
This limitation confines the practical principal number to $n=60$ in current implementations \cite{Lev19}.
In contrast, a continuous 2$\pi$ rotation to the $n=100$ Rydberg state successfully retrieves the qubit basis within the error margins associated with spontaneous emission.
 Designing gate operations that utilize a continuous laser pulse thus enables excitation to large $n$, benefiting from an enhanced interaction-to-loss ratio that scales with $n^{14}$.

The sensitivity of the Rydberg state to stray fields increases with larger principal numbers, as the polarizability scales with $n^7$. 
Continuous pulses offer laser protection through a mechanism akin to the dynamic decoupling approach in NMR, which isolates the system from slowly varying noise by consistently rotating the system perpendicularly to the noise direction. This results in the errors averaging to zero over time.
Our approach emphasizes operating Rydberg gates with a single continuous pulse, facilitating a more efficient and reliable retrieval of the qubit basis, even for highly excited Rydberg atoms with large principal numbers.

 \begin{figure}
\centering
\raggedleft
\scalebox{0.51}{\includegraphics*{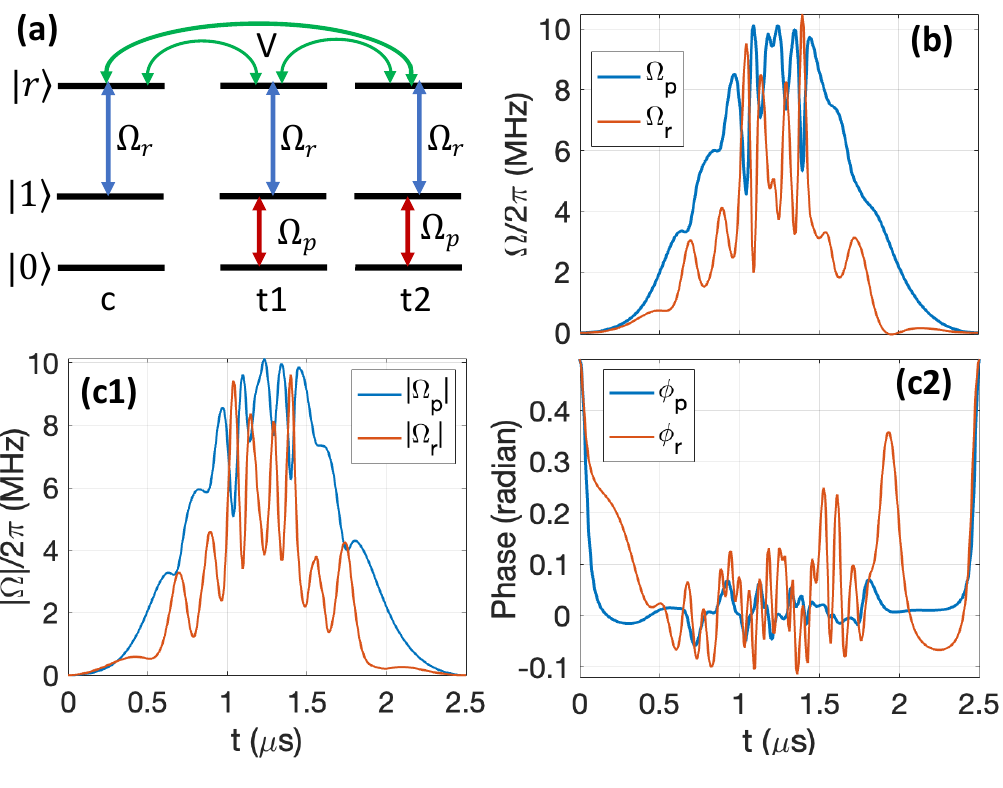}}
\caption{{\bf Fredkin (controlled-swap) gate} (a) Level scheme.  The qubit basis is the Hyperfine states of Cs, and $|r\rangle$ is the highly excited Rydberg state $|100S_{1/2}\rangle$.  Target qubit rotation is achieved via the $\Omega_p$ laser. 
To make the swap conditional,  all control and target qubits in state $|1\rangle$ are coupled to the Rydberg level. The three atoms are within the blockade region $\Omega_r \ll V$ in a triangular spatial geometry.
(b) Optimal Rabi frequencies that lead to Fredkin gate with 0.0078 infidelity. Here only amplitudes are modulated while the lasers' phase are constant.
 Alternatively (c) optimizes both (c1) amplitude and (c2) phase of the Rabi frequencies allowing for further reduction of the infidelity to 0.0026. Reported infidelities include spontaneous emission as well as rotation errors. }\label{Fig1}
\end{figure}

\section{Results}

{\bf Multi-qubit Controlled-Rotation - }
In the first category, controlled rotation gates, we apply the Rydberg exciting lasers to both control and target qubits, while the rotation pulses are exclusively applied to the target qubits. This is illustrated in Fig.~\ref{Fig1}a, the level scheme of  {\it Fredkin}  (C-SWAP) gate.  
The qubit basis consists of long-lived hyperfine ground states of $^{133}$Cs, $\ket{0}=\ket{6S_{1/2}, F=3}$ and $\ket{1}=\ket{6S_{1/2}, F=4}$, which can be coherently coupled in target atoms using a microwave or optical Raman transition with Rabi frequency $\Omega_p(t) = |\Omega_p(t)| \exp(i \phi_p(t))$. 
The qubit states $\ket{1}$ of both control and target qubits are resonantly coupled with Rabi frequency $\Omega_r(t)=|\Omega_r(t)|\exp(i \phi_r(t))$ to the Rydberg state $\ket{r}$ via a one- or two-photon process.
In a triangular lattice, equal van der Waals interactions $V \gg \Omega_r$ between each pair of atoms shift the energy of $\ket{r_i r_j}$ out of resonance with the laser.
The Hamiltonian of the control and target qubits is given by:
 \begin{eqnarray}
 \label{Eq1}
 &&H=\sum_{i= t1,t2} \frac{\Omega_p(t)}{2} (\ket{0_i}\bra{1_i}+h.c.)+ \\ \nonumber
 &&\sum_{j=c,t1,t2}[\frac{\Omega_r(t)}{2}  (\ket{r_j}\bra{1_j}+h.c.)+i\gamma_r\ket{r_j}\bra{r_j}]+\sum_{k<l} V \ket{r_kr_l}\bra{r_kr_l}
\end{eqnarray}
where $\gamma_r$ represents the spontaneous emission from the Rydberg level \cite{Bet09}. 
 \begin{figure*}
\centering
\raggedleft
\scalebox{0.54}{\includegraphics*{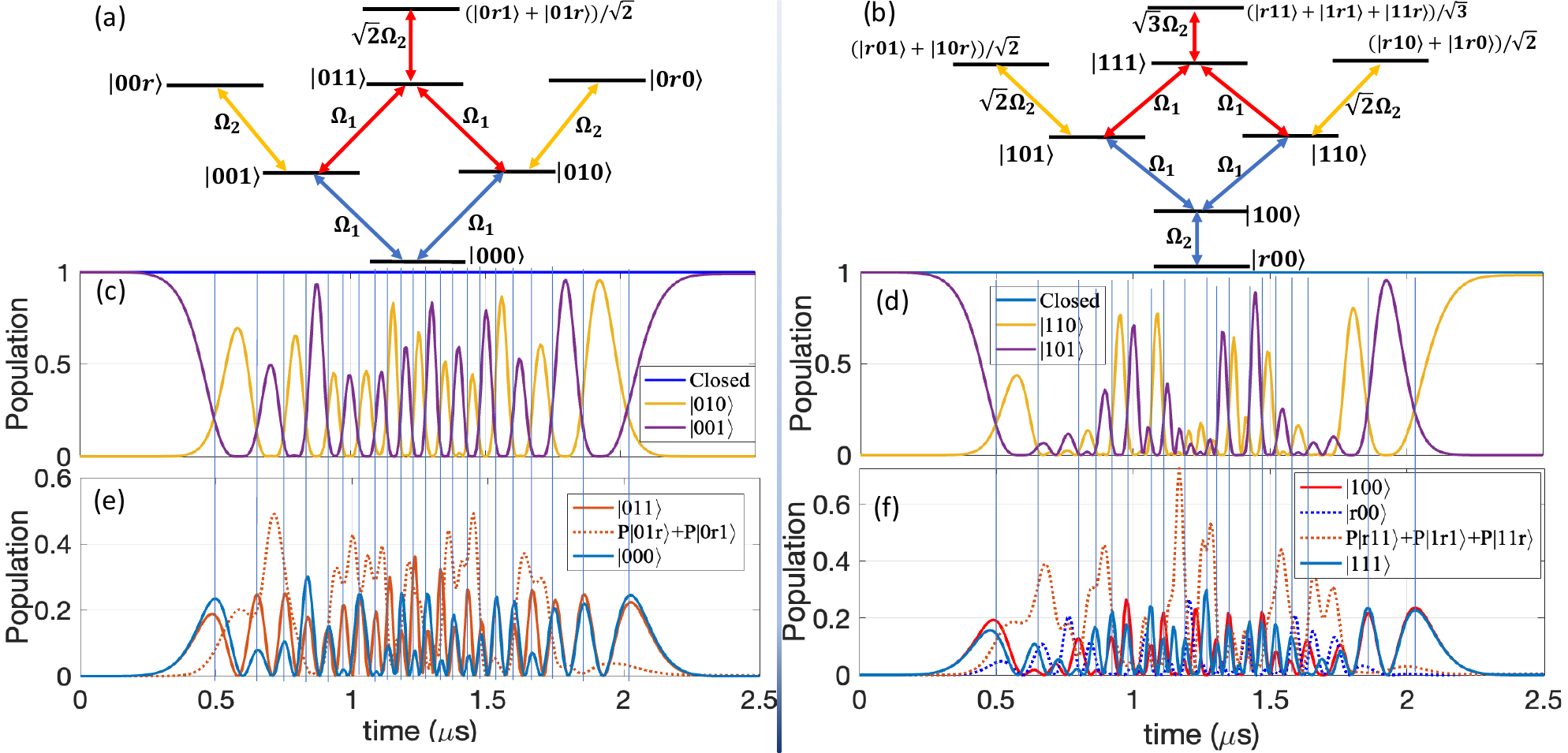}}
\caption{ {\bf Controlled-swap Mechanism - } The two closed subspaces governing the dynamics for (a) $|0_c\rangle$ and (b) $|1_c\rangle$ are illustrated. The two Lambda transitions, shown in red and blue, mediate the swapping operations, controlled by quantum interference \cite{Kha24,Bou02,Sor00} through the Rydberg transition \cite{Kha24}. Under the blockade effect, the enhanced Rabi oscillations $\sqrt{m}\Omega_2$ control the swapping rate in the Lambda transitions, enabling controlled swap operations. The orange side transition regulates the acquired phase.  
(c, d) Display the conditional swapping process and the total population within the closed subspaces defined in (a, b), as shown by the blue lines.  
(e, f) Show the population dynamics of intermediate states in the upper and lower Lambda transitions, peaking at the moments of population swap observed in (c, d). 
}\label{FigClosedSubSpace}
\end{figure*}
In the next step, the QOC precisely tailors the temporal profiles of the Rabi frequencies in the presence of strong Rydberg interactions to generate the desired Fredkin outcome states for all input qubit configurations. Figures~\ref{Fig1}b and \ref{Fig1}c illustrate the controlling profile for cases with constant and variable laser phases, respectively.

The mechanism of the conditional swap can be better understood in the collective basis representation, as shown in Fig.~\ref{FigClosedSubSpace}. The system evolves within one of two closed subspaces, depending on whether the control qubit is in the $|0_c\rangle$ or $|1_c\rangle$ state, as depicted in Fig.~\ref{FigClosedSubSpace}(a, b). To further clarify this, we show the population dynamics within each subspace in Fig.~\ref{FigClosedSubSpace}(c, d), where the blue lines highlight that the subspaces remain distinct throughout the evolution. The evolution of other qubit configurations is presented in the supplementary material. 

The operation dynamics result from a complex interplay of effects orchestrated by the quantum optimal control (QOC) pulses. Broadly speaking, the swap between $\ket{01}$ and $\ket{10}$ is mediated by upper and lower Lambda transitions, represented in Fig.~\ref{FigClosedSubSpace}(a, b) with red and blue color coding, respectively. These transitions are controlled by quantum interference \cite{Kha24,Bou02,Sor00} that is modulated by the Rydberg interaction via the relative ratio $\sqrt{m}\Omega_2/\Omega_1$ \cite{Kha24}, where $m \in \{0, 1, 2, 3\}$ is the number of atoms in resonance with the Rydberg exciting laser, i.e. in $\ket{1}$ state, which translates to the enhanced coupling arising from the blockade effect.  

For the QOC pulses shown in Fig.~\ref{Fig1}b, where the condition $\Omega_2 < \Omega_1$ holds for most of the pulse sequence, the intermediate states (e.g., $\ket{111}$ and $\ket{100}$) in both the upper and lower Lambda transitions become significantly populated at key moments during the swap process. This behavior is evident in Fig.~\ref{FigClosedSubSpace}(c–f), which highlights the population peaks in these states during the swap.  
Additionally, side Rydberg transitions, indicated by the orange arrows in Fig.~\ref{FigClosedSubSpace}, contribute to the dynamics by rotating the population into the Rydberg states. This process fine-tunes the phase parameters, thereby enhancing control over the operation's dynamics.

\begin{figure}
\centering
\scalebox{0.33}{\includegraphics*{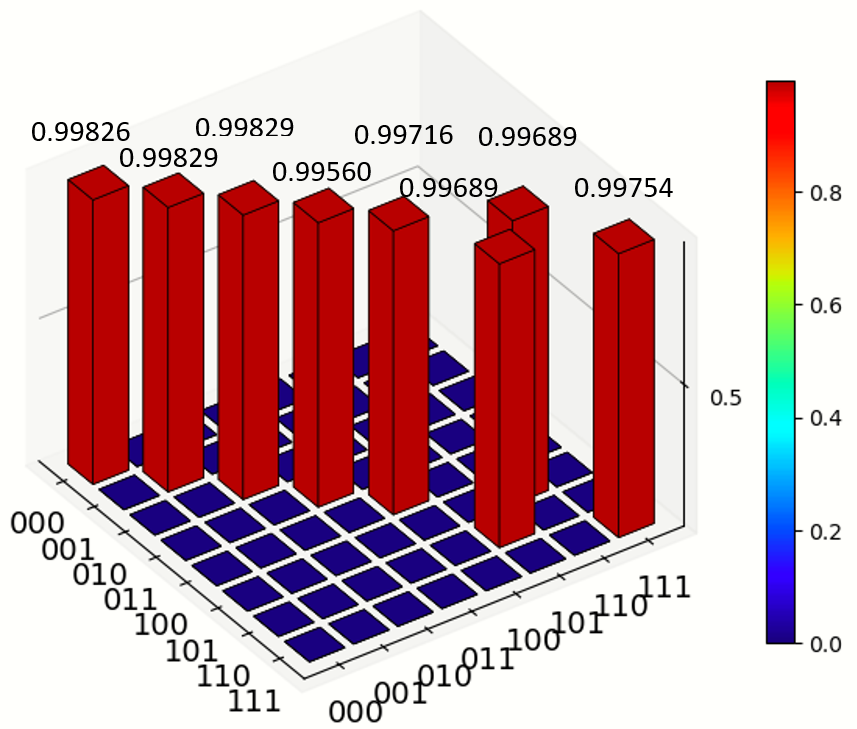}}
\caption{ Truth tables of the Fredkin gate operated with the set of parameters of Fig 1c.
}\label{Fig2}
\end{figure}

The performance of the Fredkin gate under the laser pulses of Fig.~\ref{Fig1} c is visualized in the truth table of Fig.~\ref{Fig2}. The 8×8 truth table is obtained by calculating the transition probability between initial and final qubit configurations $|\bra{\phi_f}U_F\ket{\phi_i}|^2$ under the Fredkin operation $U_F$. The time evolution of individual qubit configurations is presented in the supplementary material. 
In quantifying the gate performance, we obtain high fidelity operations of 99.2\% and 99.74\% for the set of optimized pulses presented in Fig.~\ref{Fig1}b,c respectively.

 The main source of decoherence in atomic processors is the short lifetime of Rydberg states. In the quantum optimal control scheme, the cost function $J$ minimizes the Rydberg population, leading to higher fidelities, see Methods.
To compare this method with the conventional Rydberg schemes, we define the time-integrated probability to be in the Rydberg state $\bar{T}_r= \int P_r(t) dt$ where $P_r(t)$ is the Rydberg population averaged over all qubit configurations at time $t$. For the parameter set of Fig.~\ref{Fig1}c the time-integrated Rydberg population over the gate operation is $\bar{T}_r=0.2\mu$s. 
In an alternative circuit model, the implementation of the Fredkin gate requires 8 C-NOT gates, see Supp. Applying a similar Gaussian Rydberg exciting pulse for making the 8 C-NOTs, the total time-integrated Rydberg population would be $\bar{T}_r=2.4\mu$s. Hence, the quantum optimal control scheme significantly suppresses the Rydberg population. 

\begin{figure}
\centering
\raggedleft
\scalebox{0.42}{\includegraphics*{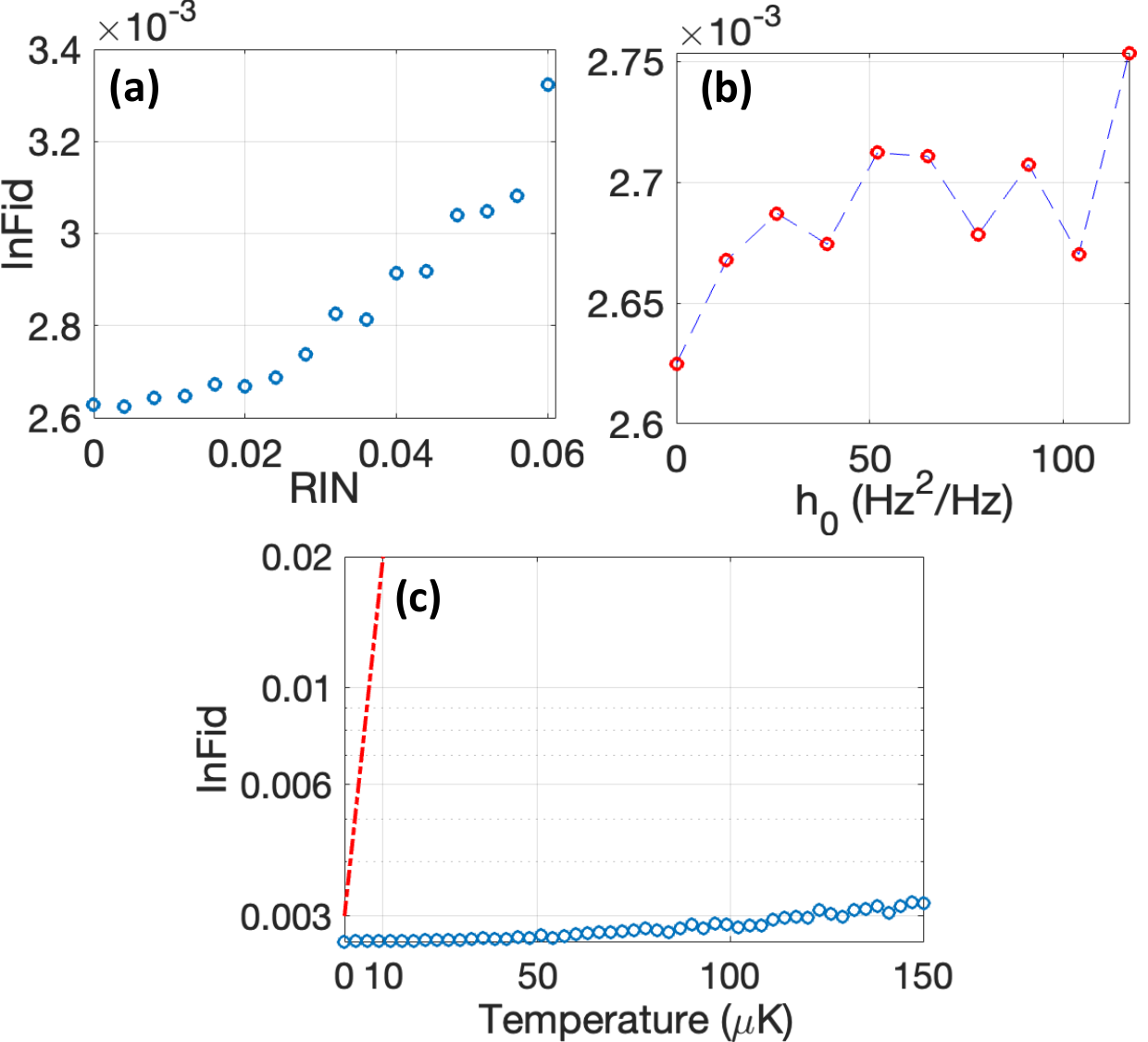}}
\caption{ Robustness against fluctuations. (a) The gate infidelity as a function of the laser intensity fluctuations for the set of optimized pulses in 
fig.~1c. The data are generated by Gaussian random sampling with different half-width at half maximum of relative intensity noise (RIN).   (b) Effects of the laser noise of Eq. \ref{Eq2} on the fidelity. The parameters are extracted from \cite{Jia23}, see the text. The effects of the servo bumps are negligible. The sensitivity to the white noise is plotted in (b).  (c) Doppler dephasing. The scattered points show infidelity due to the Doppler effect as a function of atomic temperature. Each point is averaged over 50 runs of gate simulation with the pulses of Fig.~\ref{Fig1}c. The results are compared with the Doppler sensitivity of the Rydberg antiblockade Fredkin gate (dotted-dashed line) extracted from \cite{Wu21}.
}\label{Fig3}
\end{figure}

Moreover, we investigate the robustness of the proposed scheme against laser fluctuations and Doppler-induced dephasing. Amplitude fluctuations, which can originate from spatial inhomogeneities or power drifts in the laser field, lead to uncertainty in the Rabi frequency. To mitigate spatial inhomogeneity, a super-Gaussian laser profile with a flat-top intensity distribution could be employed \cite{Gil16}. The impact of power drifts is analyzed by introducing Gaussian-distributed random fluctuations in the Rabi frequency at each time step. Figure \ref{Fig3}a illustrates the infidelity as a function of the half-width at half-maximum (HWHM) of the relative intensity noise (RIN). In 1040nm Ti:Sa lasers, the RIN has been successfully suppressed to below 0.075\% across a broad bandwidth from 10 Hz to 10 MHz \cite{TISa}. Consequently, the effect of laser intensity noise on the system is negligible.

  The motion of atoms can cause a Doppler shift, leading to an effective detuning of the resonant Rydberg exciting laser. 
    The three Doppler-induced detunings experienced by atoms in each time step are random variables, following a Gaussian probability distribution with a mean of $\bar{\delta} = 0$ and a standard deviation $\sigma_{\delta} = k_{\text{eff}}v_{\text{rms}}$. 
  Here, $k_{\text{eff}}$ represents the effective wavevector of the lasers interacting with the atoms, and $v_{\text{rms}} = \sqrt{k_B T_a / M}$ is the atomic root-mean-square velocity, where $k_B$ is the Boltzmann constant, $T_a$ is the atomic temperature, and $M$ is the atomic mass. 
In the Cs atoms, the two-photon Rydberg pumping via the intermediate state $\ket{7P_{1/2}}$, counterpropagating laser fields with 1040nm and 459.4nm wavelengths are considered.
The results indicate a notable resilience to the Doppler effect compared to the anti-blockade Rydberg Fredkin gate \cite{Wu21}. The operation fidelity that requires evaporative cooling down to 10$\mu$K in the anti-blockade Fredkin gate \cite{Wu21} could be performed with laser cooling to 150$\mu$K in our scheme, see Fig.~\ref{Fig3}b. This makes the experimental realization significantly simpler.

Laser stabilization is typically achieved by locking the laser to a narrow-linewidth reference cavity \cite{Dre83}, with the error signal processed through a feedback system or servo loop \cite{Rie04}. Due to the finite bandwidth of the servo loop, distinct peaks—referred to as servo bumps—emerge in the power spectral density $S(f)$, usually centered around 1 MHz on either side of the main peak. Experimental observations for the 1040nm Ti:Sa laser \cite{Jia23} indicate that these servo bumps follow an approximately Gaussian profile. The overall noise spectrum can be modeled as a combination of Gaussian-shaped servo bumps and white noise, expressed as \cite{Jia23}:
\begin{equation} \label{Eq2}
S_{\delta \nu} = h_0 + \frac{s_g f_g^2}{\sqrt{8\pi} \sigma_g} \left( e^{-\frac{(f - f_g)^2}{2 \sigma_g^2}} + e^{-\frac{(f + f_g)^2}{2 \sigma_g^2}} \right)
\end{equation}
Here, $h_0$ denotes the amplitude of the white noise power spectral density, $\sigma_g$ is the width of the Gaussian servo bump, and $f_g$ is its central frequency. The full width at half maximum (FWHM) of the Gaussian bump is given by $ \text{FWHM} = \sqrt{\ln(4)} \, \sigma_g $. 

For the two-photon Rydberg excitation in $^{133}$Cs, involving an intermediate state of $\ket{7P}$, the applied lasers operate at wavelengths of 1040 nm and 459 nm. The 459 nm laser is frequency-locked to an atomic transition, while the 1040 nm laser is stabilized by locking to a reference cavity. The noise characteristics of the 1040 nm laser have been determined in previous studies \cite{Jia23, Gra22}, which include a combination of white noise and two Gaussian-shaped servo bumps. The noise parameters are given by \cite{Jia23}: $ h_0 = 13 \, \text{Hz}^2/\text{Hz} $ for the white noise, and for the servo bumps $ h_{g1} = 25 \, \text{Hz}^2/\text{Hz}, \sigma_{g1} = 18 \, \text{kHz}, f_{g1} = 130 \, \text{kHz} $, and $ h_{g2} = 2.0 \times 10^3 \, \text{Hz}^2/\text{Hz}, \sigma_{g2} = 1.5 \, \text{kHz}, f_{g2} = 234 \, \text{kHz} $. The effects of this noise on fidelity are negligible. Even an order of magnesium larger servo bump noise does not change the Fidelity magnitude. Effects of background white noise are plotted in Fig.~\ref{Fig3}c. 

A real fluctuating Gaussian process with zero mean and variance $\sigma^2$ can generally be represented as a Fourier series \cite{Jia23}:
\begin{equation}\label{Eq3}
\delta \nu(t) = \sum_{j=1}^{\infty} \delta \nu_j \sin(2\pi f_j t + \phi_j)
\end{equation}
where $ f_j = j \Delta f $, $ \delta \nu_j = -2\sqrt{S_{\delta\nu}(f_j) \Delta f} $, and the random phases $\phi_j$ are drawn from a uniform distribution over $[0, 2\pi]$. This frequency noise induces detuning on the Rydberg transition. The fidelity of the resulting process averaged over 50 runs as a function of the noise parameters, is shown in Fig.~\ref{Fig3}c.

Finally, the interaction variations due to spatial uncertainty have a negligible effect on fidelity, as the Rabi frequency is already three orders of magnitude smaller than the interaction strength. With a 3 µm lattice spacing and single-site confinement to a half-width at half-maximum (HWHM) of 30 nm—achievable in optical lattices—the mean deviation in interaction strength is only 6\%, which fully preserves the blockade effect. This offers a significant advantage over the antiblockade Fredkin gate, where even a 1 nm variation in interatomic distance drastically reduces fidelity, as shown in Fig. 7 of \cite{Wu21}.

\section*{Discussion}

In this work, we have deployed quantum optimal control (QOC) as an efficient approach for implementing high-fidelity operations in Rydberg-based quantum processors. By leveraging continuous, optimized laser pulses, we have successfully reduced the operational complexity and enhanced the robustness of critical multi-qubit gates such as the Fredkin with 99.74\% fidelity, surpassing the performance of previous approaches that rely on conventional gate decompositions or sequences of pulses. Our computational optimization framework, built on the Krotov method, identifies time-dependent laser parameters, minimizing population in Rydberg states and significantly reducing decoherence effects.

The combination of quantum optimal control (QOC) with the unique properties of the Rydberg platform presents a powerful strategy for overcoming one of the major challenges in quantum computing: scalable and fault-tolerant quantum gate operations. Conventional approaches based on anti-blockade mechanisms and stepwise laser control sequences are highly susceptible to errors caused by environmental noise, Doppler dephasing, and uncertainties in interatomic distances, all of which degrade gate fidelity. In contrast, our continuous control approach inherently mitigates these issues by maintaining robust laser protection, minimizing population in excited states during gate execution, and ensuring that the interaction-induced energy shift remains significantly larger than both Doppler broadening and the broadening caused by interatomic distance fluctuations. This strong interaction shift effectively preserves the blockade condition, preventing detrimental effects from level broadening and ensuring high-fidelity gate operations.

From a computational science perspective, the application of QOC to multi-qubit operations highlights the potential of machine-learning-assisted optimization techniques and advanced control algorithms to push the boundaries of current quantum hardware. The use of optimized pulse sequences not only improves gate fidelity but also reduces the physical resources required to perform quantum operations, providing a pathway to more resource-efficient quantum processors.

 The ability to execute high-fidelity multi-qubit gates with fewer operational steps reduces the accumulated error, making our approach particularly relevant for near-term quantum algorithms. The computational framework presented here can be extended to other applications \cite{Kha22,Kha25}. One specific example is making large Schr\"odinger cat states \cite{Kha16,Kha18,Khazali16,Kha24cat} and pattern formation \cite{Kha21,KhaShi24} in the presence of strong Rydberg dressing \cite{Kha24cat}. In Rydberg dressing, while going toward resonance enhances the interaction-to-loss ratio desired for making large cat states, the rising number of multi-component non-linear terms suppresses the fidelity of the target state \cite{Kha24cat}. The optimal control of the dressing strength could be used to obtain large-scale, high-fidelity cat states and allow the formation of new types of quantum matter, broadening QOC applicability across various quantum architectures.

 Future efforts should explore how these optimized gates can be integrated into larger quantum circuits, focusing on the effects of noise and decoherence as the system scales. Additionally, the development of real-time adaptive control schemes using QOC could further enhance gate performance by dynamically adjusting laser pulses in response to hardware-specific imperfections.

\section*{Methods}
 {\bf Krotov's Method for Quantum Optimal Control - }
A quantum optimal control problem involves finding control fields (external forces or pulses) that steer a quantum system from an initial state to a desired target state while minimizing certain cost functions. This approach is fundamental in addressing various challenges in quantum computing, such as implementing quantum gates and state transfer.
Among the numerous methods proposed for quantum optimal control, Krotov and GRAPE (Gradient Ascent Pulse Engineering) are two commonly used gradient-based optimization algorithms  \cite{Goerz2019Krotov,Morzhin2019}.
Krotov's method, initially proposed outside quantum control theory by Krotov and Feldman (1978, 1983), was later adapted for quantum systems by Tannor, Kazakov, and Orlov (1992). This method guarantees monotonic convergence for near-continuous control fields, making it particularly useful for exploring the limits of controllability in physical systems.

The time evolution of a quantum state in a closed quantum system is governed by the Schrödinger equation:
\begin{equation}
i\hbar \frac{\partial |\Psi(t)\rangle}{\partial t} = H(t) |\Psi(t)\rangle.
\end{equation}
To apply the quantum optimization we need to linearize the Hamiltonian in terms of control parameters $\chi$
\begin{equation}
H(t) = H_0 + \sum_{i=1}^{N} \chi_i(t) H_i.
\end{equation}
For example, in the case of the variable phase in Fig.~\ref{Fig1}c, the Hamiltonian of Eq.~\ref{Eq1} could be linearized by decomposing the Rabi frequencies into the real and imaginary components as follows:
\begin{equation}
\frac{\Omega^{(re)}(t)}{2}\ket{\alpha}\bra{\beta}+i \frac{\Omega^{(im)}(t)}{2} \ket{\alpha}\bra{\beta}+h.c.
\end{equation}
and optimizing the real and imaginary parts separately.

In the context of implementing quantum gates, the objective is to find the set of controls $\{\chi_i(t)\}$ such that at $t = T$, in the computational basis, we obtain the desired set of states $\{\Psi_j(T)\}$. This can be achieved by formulating the problem using the variational principle, which involves finding functions that minimize a functional dependent on those functions.

For optimizing gate operations, we define a functional $J[\{\chi_i(t)\}, \{\Psi_j(t)\}]$, where the condition for $J$ to be minimized is that its variations with respect to $\chi$ and $\Psi$ are zero:
\begin{equation}
\nabla J_{\chi_i, \Psi_j} = 0.
\end{equation}
The functional is given by
\begin{equation}
J[\{\chi_i(t)\}, \{\Psi_j(t)\}] =  J_T\{\Psi_j(T)\}+\sum_i \int_0^{T} g_a(\chi_i(t)) \, dt ,
\end{equation}
where $J_T$ is the gate infidelity at the end of the operation.
The running cost on the control fields is denoted as $g_a$, which quantifies the change applied to the control fields in each iteration. 
\begin{equation}
g_a \left( \chi_l^{(i)}(t) \right) = \frac{\lambda_{a,l}}{S^l(t)} (  \chi_l^{(i)}(t) - \chi_l^{(i-1)}(t))^2,
\end{equation}
where $\lambda_{a,l} > 0$ is the inverse “step width,” $S_l(t) \in [0, 1]$ is the “update shape” function.
Minimizing this term indicates approaching the final optimal pulse.
A comprehensive review of the mathematical foundations and applications of Krotov's method can be found in the literature  \cite{Goerz2019Krotov,Morzhin2019}.

In this work, we emphasize the critical importance of optimally selecting the parameter $\lambda$ in our optimization problem. Identifying the optimal value of $\lambda$ is a non-trivial task that typically requires iterative testing and refinement. If $\lambda$ is set too high, the optimization process may fail to converge, whereas if $\lambda$ is too low, the resulting control pulses will exhibit rapid, non-smooth variations, rendering them impractical for experimental implementation. Unfortunately, there is no definitive method for determining the ideal $\lambda$, necessitating a try-and-error approach.

Additionally, the initial shape of the pulse, as well as the starting values for its intensity and phase, play crucial roles in achieving high fidelity. Our results demonstrate that by carefully finding the optimum parameters, we can achieve pulses with acceptable levels of smoothness, making them viable for experimental use. Furthermore, our optimized pulses achieve a level of fidelity for the Fredkin gate that, to our knowledge, is unprecedented.

   {\bf Data availability}  All the data that support the findings of this study are included in this article and Supplement 1. Any additional data is available from the corresponding author upon reasonable request.

  {\bf Code availability}
All relevant code supporting the document is available upon request.

{\bf Supplemental document} See Supplement 1 for supporting information and further discussions.
  
 {\bf  Author contributions} All authors contributed equally to the conceptualization,  analysis, and writing of this work.

 {\bf  Competing interest}
The authors declare no competing interests.

\clearpage
\begin{widetext}
\begin{center}
{\bf \Large Supplementary Information}
\end{center}
\end{widetext}

\twocolumngrid

\setcounter{equation}{0}
\setcounter{figure}{0}
\setcounter{table}{0}
\setcounter{section}{0}
\renewcommand{\thesection}{S\arabic{section}}
\renewcommand{\theequation}{S\arabic{equation}}
\renewcommand{\thefigure}{S\arabic{figure}}
\renewcommand{\thetable}{S\arabic{table}}

\section{ Multi-qubit Controlled-Phase operation }
\begin{figure}
\centering
\raggedleft
\scalebox{0.58}{\includegraphics*{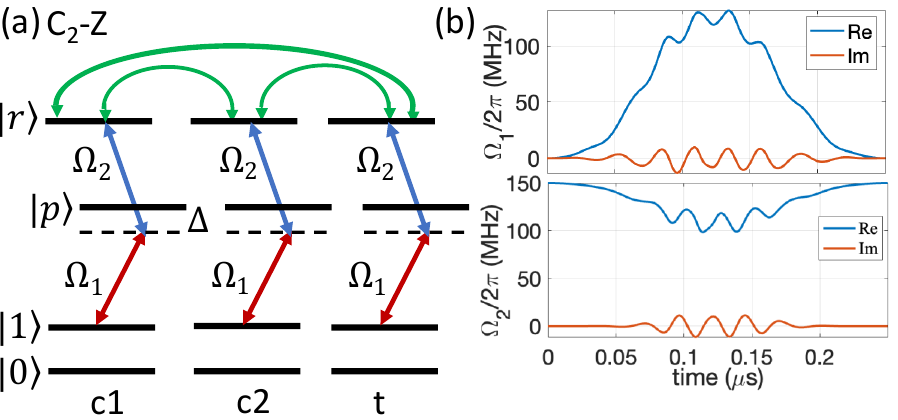}}
\caption{{\bf Controlled-Phase multi-qubit gate} (a) The level scheme for performing a   C$_2$-Z gate in the Cs atomic lattice. To exploit maximum control, we control the intensity and phase of 459.4 and 1040 nm lasers acting on a 2-photon Rydberg excitation. (b) The optimal profile of Rabi frequencies with real and imaginary parts is plotted by red and blue lines, respectively.  The simulation considers exciting atoms to the Rydberg state $|r\rangle=|70S_{1/2}\rangle$ via the intermediate state $|p\rangle=|7P_{1/2}\rangle$ with laser detuning of 1GHz. Here a triangular lattice with 3$\mu$m interatomic distance is considered. 
}\label{FigS1}
\end{figure}

In the main text, we demonstrated the use of optimal control for the two-qubit controlled swap gate. In a second category of multi-qubit controlled phase operations, we exclusively apply the Rydberg exciting pulses and do not include qubit rotation pulses. As an example, we study the C$_2$-Z$=2\ket{{111}}\bra{{111}}-\mathbb{I}$ gate, which could easily be converted to Toffoli by sandwiching the target qubit with Hadamards. This procedure is not limited to certain qubit numbers and could be extended to more complicated algorithms \cite{Pet16}, with an example of C$_3$Z provided in Fig.~S3.

Figure \ref{FigS1}a presents the level scheme. 
The {\it gate operation} is performed by exciting the $\ket{1}$ qubit state of controls and target qubits to the Rydberg level via a global pulse.  
 The Rydberg excitation is applied by two-photon excitation with $\Omega_{1}$ and $\Omega_{2}$ being the Rabi frequencies and the lasers are red detuned from the  $\ket{p}=\ket{7P_{1/2}}$ intermediate state by $\Delta$ and tuned to $\ket{r}=\ket{70S_{1/2}}$ Rydberg state. The effective gate Hamiltonian acting on the two spin-waves is
\begin{eqnarray}
\label{Eq_H}\nonumber
&&H=\sum_{i\in \{c_1,c_2,t\}} \frac{\Omega_1(t)}{2}\sigma^{(i)}_{1p}+\frac{\Omega_2(t)}{2}\sigma^{(i)}_{rp}+h.c. +\Delta \sigma^{(i)}_{pp}\quad  \\ 
&&+\sum_{i<j}V({\bf x_i}-{\bf x_j})\sigma^{(i)}_{rr}\sigma^{(j)}_{rr}
\end{eqnarray}
where $\sigma^{(i)}_{lm}=\ket{l}\bra{m}$ is the transition operator of the $i^{th}$ qubit.  
 
 \begin{figure} 
\centering
\scalebox{0.29}{\includegraphics*{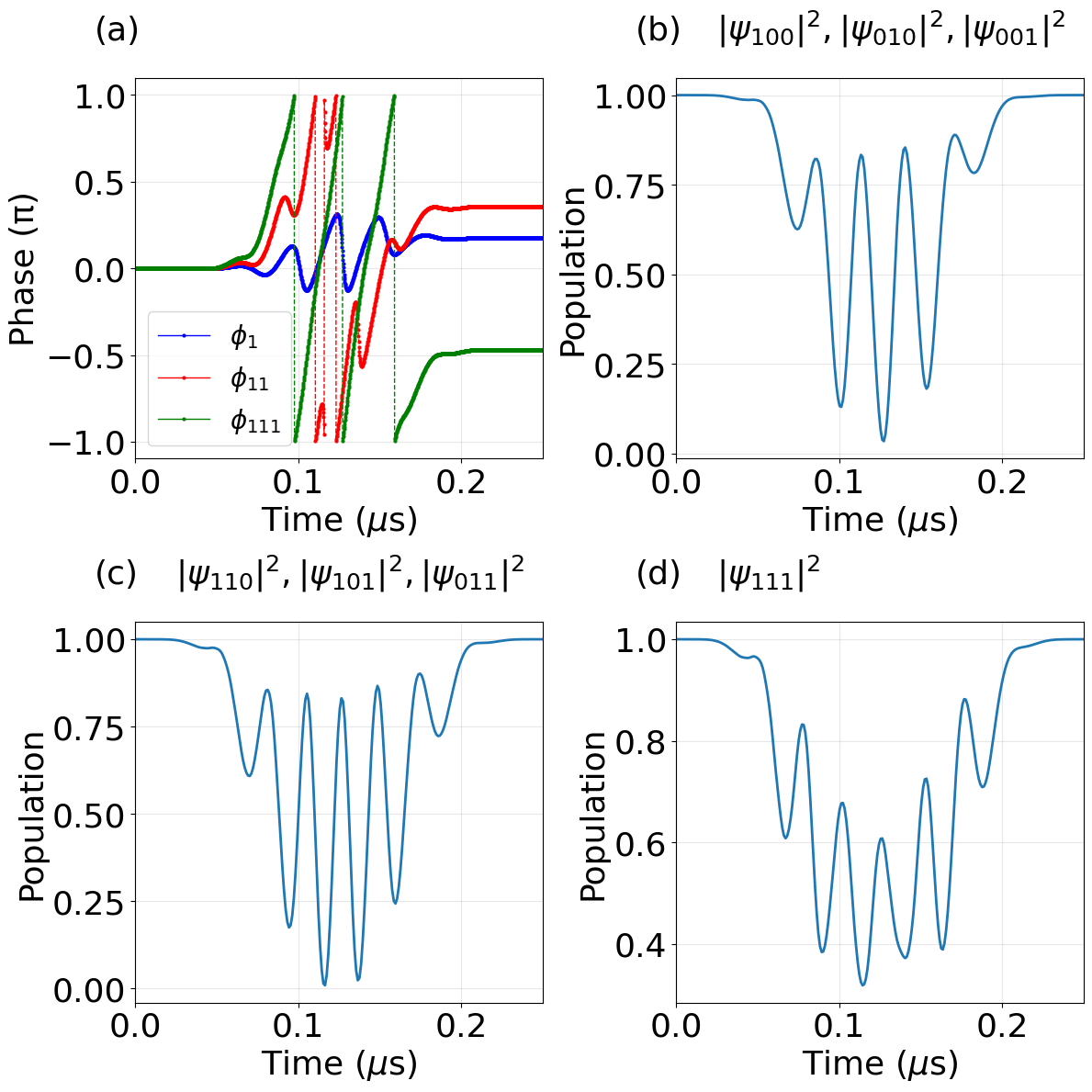}}
\caption{Time evolution of the population and phase of different qubit configurations, in the C$_2$-Z gate under the optimized pulse shapes presented in figure S1. In C$_2$-Z gate, populations return to the initial qubit basis while depending on the number of qubits in $|1\rangle$ basis the acquired phase would be $\phi_{111}=\pi+3\phi_{1}$ and $\phi_{11}=2\phi_{1}$ where $\phi_{111}$, $\phi_{11}$ and $\phi_{1}$ are the final phase that is obtained by qubit configurations with three, two, and one atoms in $|1\rangle$ qubit state.  }\label{FigS2}
\end{figure}

 \begin{figure} 
\centering
\raggedleft
\scalebox{0.8}{\includegraphics*{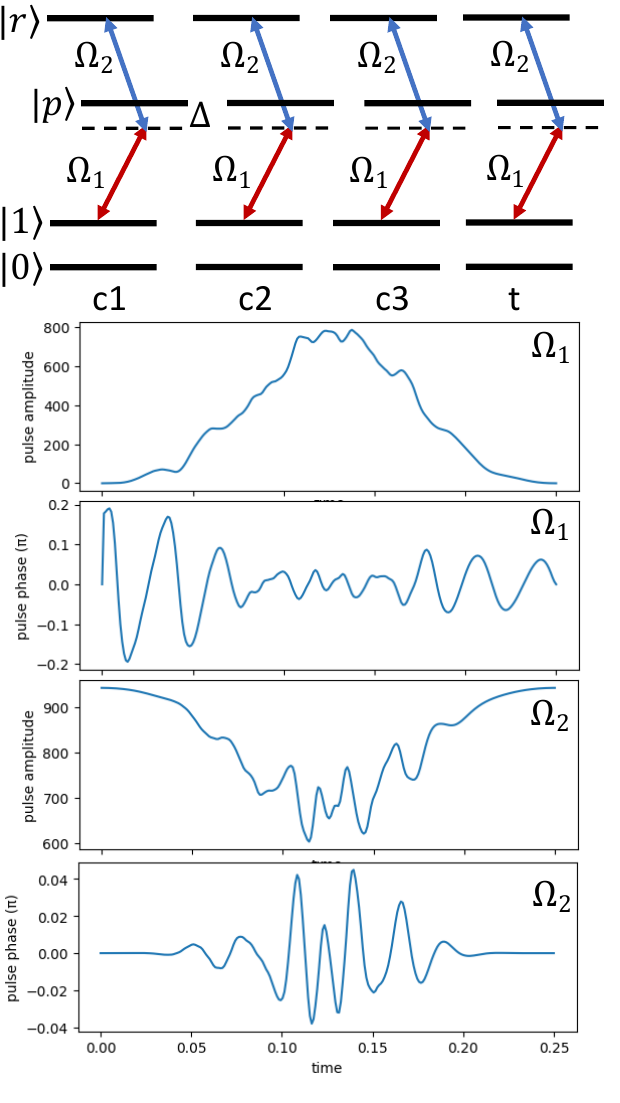}}
\caption{Optimized pulses for single step implementation of C$_3$-Z gate.}\label{FigS3}
\end{figure}

 To apply the desired C$_2$-Z  gate with a single continuous global pulse, we deploy the Krotov optimization technique to engineer the time-varying amplitudes and phases of the lasers used for two-photon Rydberg excitation.
 Figure \ref{FigS1}b shows the time evolution of the Rabi frequencies that retrieve the computational basis for all qubit configurations with the desired conditional phase arrangement.
 For the qubit configurations with one atom in $\ket{1}$ state i.e. $\ket{100}$, $\ket{010}$ and $\ket{001}$ the gate operation supplies a phase of $\phi_{1}$ and retrieves the qubit basis. 
 In the case of two or three qubits being in the $\ket{1}$ state, the presence of atoms within the blockade volume prevents more than one Rydberg excitation. The effective operation would then be in a two-dimensional space of all the atoms being in the qubit basis (e.g. $\ket{011}$) and a symmetric superposition of one of the atoms in $\ket{1}$ being transferred to the Rydberg state e.g. $\ket{w}=(\ket{01r}+\ket{0r1})/\sqrt{2}$. As a result, the system is driven by an ehanced Rabi frequency (e.g. $\sqrt{2}\Omega_{\text{eff}}$) and experiences a modified differential light-shift for the transition.  
 Hence,  distinguished qubit configurations would follow different geometric paths over the Bloch sphere. 
 To perform the C$_2$-Z gate, the optimal control technique is applied on laser parameters to retrieve the qubit basis for all qubit configurations and to obtain the desired phase arrangement $\phi_{111}=\pi+3\phi_{1}$ and $\phi_{11}=2\phi_{1}$ where $\phi_{111}$, $\phi_{11}$ and $\phi_{1}$ are the final phase that is obtained by qubit configurations with three, two, and one atoms in $|1\rangle$ qubit state.
 See Fig.~\ref{FigS2} for the population and phase evolution of different qubit configurations.

\section{Population rotations in Fredkin gate}
\label{Pop}
\begin{figure}
\centering
\raggedleft
\scalebox{0.5}{\includegraphics*{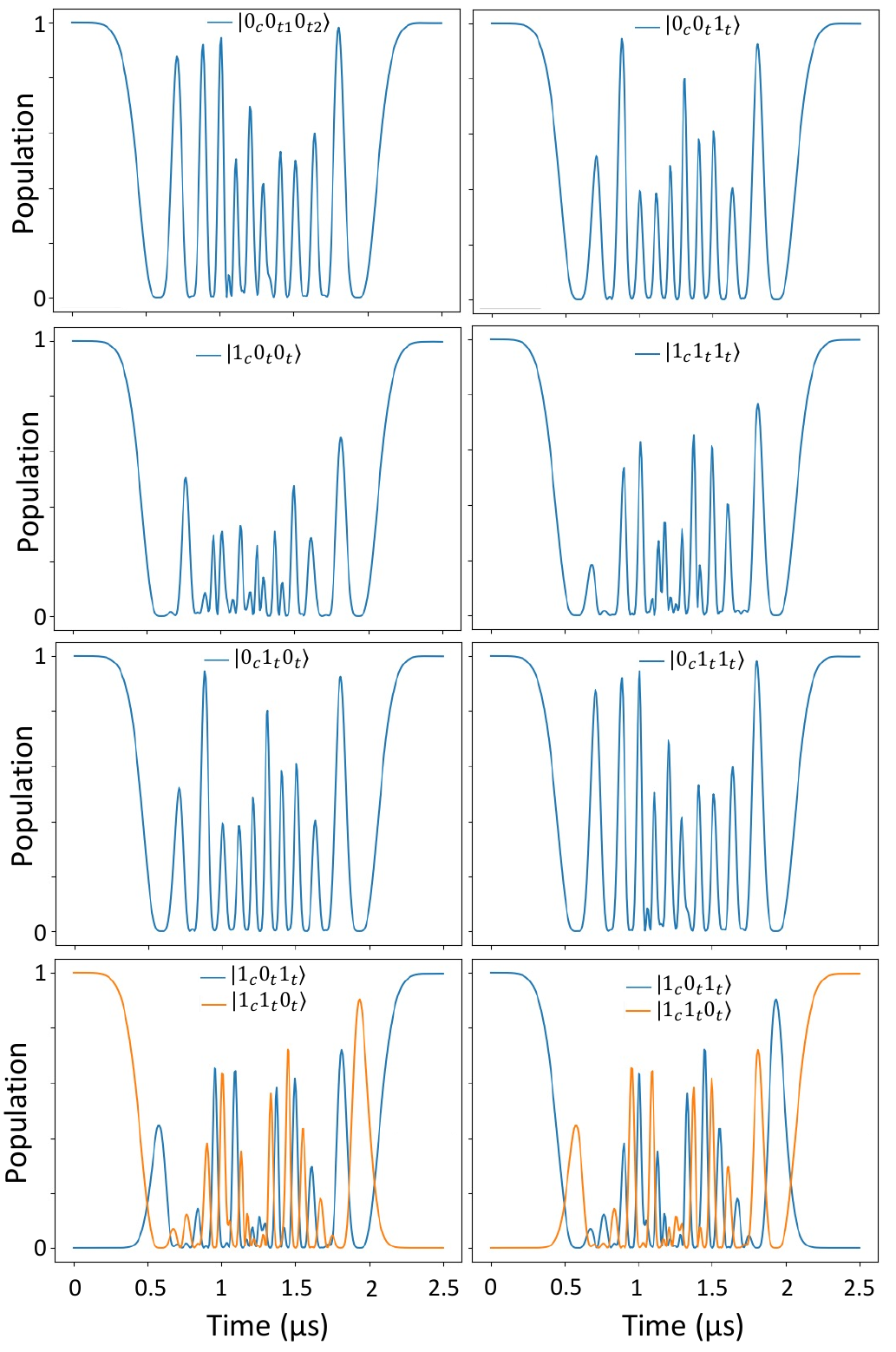}}
\caption{Time evolution of different qubit configurations, in the Fredkin gate under the optimized pulses in figure 1c of the main text.}\label{FigS4}
\end{figure}

The time evolution of different qubit configurations in the Fredkin gate under the optimized pulses presented in Figure 1c of the main text are plotted in Fig.~\ref{FigS4}.

The current scheme significantly suppresses the population of the short-lived Rydberg state. For the parameter set of Fig.~\ref{FigS1} the time-integrated Rydberg population over the gate operation is $\bar{T}_r=0.018\mu$s, resulting in 0.0001 spontaneous emission error from the $\ket{70S}$ Rydberg level without assuming the cryogenic environment. 
A similar Gaussian Rydberg exciting pulse in the conventional Rydberg Toffoli gate \cite{Ise11},  leads the time-integrated Rydberg population of  $\bar{T}_r=0.15\mu$s for C$_2$-NOT. In an alternative circuit model, the Toffoli would need 6 C-NOT gates, see Fig.~S5. Applying a similar Gaussian Rydberg exciting pulse, the total time-integrated Rydberg population would be $\bar{T}_r=0.72\mu$s. Other than the spontaneous emission, the rotation fidelity could approach unity by any desired accuracy by enhancing the number of optimization steps.

Finally, the real advantages of using the QOC appear in more complicated quantum algorithms. For example, in Toffoli (C$_k$-NOT) gate, the circuit model implementation requires 2k+2 C-NOT gates \cite{She09} and a large number of one-qubit gates. The alternative operation is a single pulse that could significantly simplify the gate and enhance the fidelity. Fig.~\ref{FigS3} represents the optimized pulse for the implementation of C$_3$-NOT.

\end{document}